\documentclass[apjl]{myemulateapj}
\usepackage{comment}
\usepackage{ifthen}

\newcommand{\ctbd}[1]{}

\newcommand{\lc}{light curve}
\newcommand{\lcs}{light curves}
\newcommand{\Lc}{Light curve}

\newcommand{\band}[1]{\ensuremath{#1}~band}

\newcommand{\kms}{\ensuremath{\rm km\,s^{-1}}}
\newcommand{\ms}{\ensuremath{\rm m\,s^{-1}}}
\newcommand{\mss}{\ensuremath{\rm m\,s^{-2}}}
\newcommand{\gcmc}{\ensuremath{\rm g\,cm^{-3}}}
\newcommand{\ergscmsq}{\ensuremath{\rm erg\,s^{-1}\,cm^{-2}}}

\newcommand{\masyr}{\ensuremath{\rm mas\,yr^{-1}}}

\newcommand{\teff}{\ensuremath{T_{\rm eff}}}

\newcommand{\vsini}{\ensuremath{v \sin{i}}}
\newcommand{\feh}{\ensuremath{\rm [Fe/H]}}

\newcommand{\vmac}{\ensuremath{v_{\rm mac}}}
\newcommand{\vmic}{\ensuremath{v_{\rm mic}}}
\newcommand{\rhk}{\ensuremath{R^{\prime}_{HK}}}
\newcommand{\logrhk}{\ensuremath{\log\rhk}}

\newcommand{\rsun}{\ensuremath{R_\sun}}
\newcommand{\msun}{\ensuremath{M_\sun}}
\newcommand{\lsun}{\ensuremath{L_\sun}}

\newcommand{\rstar}{\ensuremath{R_\star}}
\newcommand{\mstar}{\ensuremath{M_\star}}
\newcommand{\lstar}{\ensuremath{L_\star}}

\newcommand{\teffstar}{\ensuremath{T_{\rm eff\star}}}
\newcommand{\rhostar}{\ensuremath{\rho_\star}}
\newcommand{\loggstar}{\ensuremath{\log{g_{\star}}}}

\newcommand{\mearth}{\ensuremath{M_\earth}}

\newcommand{\rpl}{\ensuremath{R_{p}}}
\newcommand{\mpl}{\ensuremath{M_{p}}}

\newcommand{\rhopl}{\ensuremath{\rho_{p}}}

\newcommand{\arstar}{\ensuremath{a/\rstar}}
\newcommand{\zrstar}{\ensuremath{\zeta/\rstar}}

\newcommand{\rjup}{\ensuremath{R_{\rm J}}}
\newcommand{\mjup}{\ensuremath{M_{\rm J}}}

\newcommand{\reffigl}[1]{Figure~\ref{fig:#1}}
\newcommand{\refsecl}[1]{\mbox{Section \ref{sec:#1}}}

\newcommand{\reftabl}[1]{Table~\ref{tab:#1}}

\newcommand{\flwof}{\mbox{FLWO 1.2\,m}}

\newcommand{\flwos}{\mbox{FLWO 1.5\,m}}

\newcommand{\hatcurhtr}{HTR170-004}                                    
\newcommand{\hatcurfield}{170}                                         
\newcommand{\hatcurCCra}{\ensuremath{04^{\mathrm h}24^{\mathrm m}59.54{\mathrm s}}}                                  
\newcommand{\hatcurCCdec}{\ensuremath{+39{\arcdeg}27{\arcmin}38.3{\arcsec}}}                                 
\newcommand{\hatcurCCtwomass}{2MASS~04245952+3927382}                  
\newcommand{\hatcurCCgsc}{GSC~2883-01687}                              
\newcommand{\hatcurCCtassmv}{12.159}                                   
\newcommand{\hatcurCCtassmvshort}{12.16}                               
\newcommand{\hatcurCCtwomassKmag}{\ensuremath{9.641\pm0.019}}          
\newcommand{\hatcurCCesoKmag}{\ensuremath{9.683\pm0.020}}              
\newcommand{\hatcurCCesoJKmag}{\ensuremath{0.587\pm0.033}}             
\newcommand{\hatcurLCrprstar}{\ensuremath{0.1019\pm0.0009}}            
\newcommand{\hatcurLCbsq}{\ensuremath{0.117_{-0.046}^{+0.047}}}        
\newcommand{\hatcurLCimp}{\ensuremath{0.342_{-0.086}^{+0.060}}}        
\newcommand{\hatcurLCzeta}{\ensuremath{9.76\pm0.03}}                   
\newcommand{\hatcurLCdur}{\ensuremath{0.2285\pm0.0015}}                
\newcommand{\hatcurLCingdur}{\ensuremath{0.0237\pm0.0014}}             
\newcommand{\hatcurLCP}{\ensuremath{10.863502\pm0.000027}}             
\newcommand{\hatcurLCPshort}{\ensuremath{10.8635}}                     
\newcommand{\hatcurLCPvshort}{\ensuremath{10.9}}                       
\newcommand{\hatcurLCT}{\ensuremath{2454638.56019\pm0.00048}}          
\newcommand{\hatcurLCTA}{\ensuremath{2453649.98149\pm0.00217}}         
\newcommand{\hatcurLCTB}{\ensuremath{2454758.05871\pm0.00073}}         
\newcommand{\hatcurLCiblend}{\ensuremath{0.95\pm0.04}}                 
\newcommand{\hatcurSMEiteff}{\ensuremath{5645\pm80}}                   
\newcommand{\hatcurSMEizfeh}{\ensuremath{0.27\pm0.08}}                 
\newcommand{\hatcurSMEizfehshort}{\ensuremath{0.27}}                   
\newcommand{\hatcurSMEilogg}{\ensuremath{4.51\pm0.10}}                 
\newcommand{\hatcurSMEivsin}{\ensuremath{0.5\pm0.3}}                   
\newcommand{\hatcurSMEivmac}{\ensuremath{3.8}}                         
\newcommand{\hatcurSMEivmic}{\ensuremath{0.85}}                        
\newcommand{\hatcurSMEiiteff}{\ensuremath{5568\pm90}}                  
\newcommand{\hatcurSMEiizfeh}{\ensuremath{+0.22\pm0.08}}               
\newcommand{\hatcurSMEiizfehshort}{\ensuremath{+0.22}}                 
\newcommand{\hatcurSMEiilogg}{\ensuremath{4.39\pm0.00}}                
\newcommand{\hatcurSMEiivsin}{\ensuremath{2.0\pm0.5}}                  
\newcommand{\hatcurSMEiivmac}{\ensuremath{3.70}}                       
\newcommand{\hatcurSMEiivmic}{\ensuremath{0.85}}                       
\newcommand{\hatcurDSteff}{\ensuremath{5250\pm100}}                    
\newcommand{\hatcurDSlogg}{\ensuremath{3.5\pm0.25}}                    
\newcommand{\hatcurDSvsini}{\ensuremath{1.5\pm1.0}}                    
\newcommand{\hatcurDSgamma}{\ensuremath{+31.23\pm0.14}}                
\newcommand{\hatcurDSnumspec}{\ensuremath{6}}                          
\newcommand{\hatcurDSspan}{\ensuremath{62}}                            
\newcommand{\hatcurDSrvrms}{\ensuremath{0.33}}                         
\newcommand{\hatcurLBiz}{\ensuremath{0.2403}}                          
\newcommand{\hatcurLBiiz}{\ensuremath{0.3173}}                         
\newcommand{\hatcurLBii}{\ensuremath{0.3118}}                          
\newcommand{\hatcurLBiii}{\ensuremath{0.3113}}                         
\newcommand{\hatcurISOm}{\ensuremath{1.01\pm0.04}}                     
\newcommand{\hatcurISOmshort}{\ensuremath{1.01}}                       
\newcommand{\hatcurISOmlong}{\ensuremath{1.013\pm0.043}}               
\newcommand{\hatcurISOr}{\ensuremath{1.08\pm0.04}}                     
\newcommand{\hatcurISOrshort}{\ensuremath{1.08}}                       
\newcommand{\hatcurISOrlong}{\ensuremath{1.080\pm0.039}}               
\newcommand{\hatcurISOlogg}{\ensuremath{4.38\pm0.03}}                  
\newcommand{\hatcurISOlum}{\ensuremath{1.00\pm0.11}}                   
\newcommand{\hatcurISOmv}{\ensuremath{4.86\pm0.14}}                    
\newcommand{\hatcurISOage}{\ensuremath{6.8_{-1.6}^{+2.5}}}             
\newcommand{\hatcurISOageshort}{\ensuremath{6.8}}                      
\newcommand{\hatcurISOMK}{\ensuremath{3.17\pm0.08}}                    
\newcommand{\hatcurISOJK}{\ensuremath{0.44\pm0.02}}                    
\newcommand{\hatcurISOspec}{G5}                                        
\newcommand{\hatcurRVK}{\ensuremath{180.6\pm3.5}}                      
\newcommand{\hatcurRVk}{\ensuremath{-0.025\pm0.005}}                   
\newcommand{\hatcurRVh}{\ensuremath{-0.188\pm0.019}}                   
\newcommand{\hatcurRVgamma}{\ensuremath{-14.4\pm2.2}}                  
\newcommand{\hatcurRVjitter}{\ensuremath{7.17}}                        
\newcommand{\hatcurRVrms}{\ensuremath{7.73}}                           
\newcommand{\hatcurRVeccen}{\ensuremath{0.190\pm0.019}}                
\newcommand{\hatcurRVomega}{\ensuremath{262\pm1}}                      
\newcommand{\hatcurPPi}{\ensuremath{89.1_{-0.2}^{+0.2}}}               
\newcommand{\hatcurPPlogg}{\ensuremath{3.62\pm0.03}}                   
\newcommand{\hatcurPPar}{\ensuremath{19.16\pm0.62}}                    
\newcommand{\hatcurPParel}{\ensuremath{0.0964\pm0.0014}}               
\newcommand{\hatcurPPrho}{\ensuremath{1.96\pm0.22}}                    
\newcommand{\hatcurPPmshort}{\ensuremath{1.95}}                        
\newcommand{\hatcurPPmlong}{\ensuremath{1.946\pm0.066}}                
\newcommand{\hatcurPPrshort}{\ensuremath{1.07}}                        
\newcommand{\hatcurPPrlong}{\ensuremath{1.072\pm0.043}}                
\newcommand{\hatcurPPmrcorr}{\ensuremath{0.48}}                        
\newcommand{\hatcurPPteff}{\ensuremath{904\pm20}}                      
\newcommand{\hatcurPPtheta}{\ensuremath{0.346\pm0.015}}                
\newcommand{\hatcurPPfluxperi}{\ensuremath{2.25\pm0.205}}              
\newcommand{\hatcurPPfluxperidim}{\ensuremath{8}}                      
\newcommand{\hatcurPPfluxap}{\ensuremath{1.05\pm0.11}}                 
\newcommand{\hatcurPPfluxapdim}{\ensuremath{8}}                        
\newcommand{\hatcurPPfluxavg}{\ensuremath{1.51\pm0.137}}               
\newcommand{\hatcurPPfluxavgdim}{\ensuremath{8}}                       
\newcommand{\hatcurXsecondary}{\ensuremath{2454643.816\pm0.039}}       
\newcommand{\hatcurXsecdur}{\ensuremath{0.1607\pm0.0066}}              
\newcommand{\hatcurXsecingdur}{\ensuremath{0.0156\pm0.0009}}           
\newcommand{\hatcurXdist}{\ensuremath{190\pm8}}                        
\newcommand{\hatcurReddening}{\ensuremath{0.30\pm0.08}}                  

\newcommand{\hatcur}{HAT-P-15}
\newcommand{\hatcurb}{HAT-P-15b}

\newcommand{\hatcurSMEversion}{ii}                                       
\newcommand{\hatcurSMEteff}{\ifthenelse{\equal{\hatcurSMEversion}{i}}{\hatcurSMEiteff}{\hatcurSMEiiteff}}
\newcommand{\hatcurSMEzfeh}{\ifthenelse{\equal{\hatcurSMEversion}{i}}{\hatcurSMEizfeh}{\hatcurSMEiizfeh}}
\newcommand{\hatcurSMEzfehshort}{\ifthenelse{\equal{\hatcurSMEversion}{i}}{\hatcurSMEizfehshort}{\hatcurSMEiizfehshort}}
\newcommand{\hatcurSMElogg}{\ifthenelse{\equal{\hatcurSMEversion}{i}}{\hatcurSMEilogg}{\hatcurSMEiilogg}}
\newcommand{\hatcurSMEvsin}{\ifthenelse{\equal{\hatcurSMEversion}{i}}{\hatcurSMEivsin}{\hatcurSMEiivsin}}
\newcommand{\hatcurSMEvmac}{\ifthenelse{\equal{\hatcurSMEversion}{i}}{\hatcurSMEivmac}{\hatcurSMEiivmac}}
\newcommand{\hatcurSMEvmic}{\ifthenelse{\equal{\hatcurSMEversion}{i}}{\hatcurSMEivmic}{\hatcurSMEiivmic}}

\newboolean{emulateapj}
\setboolean{emulateapj}{true}

\newboolean{rvtablelong}
\setboolean{rvtablelong}{true}

\newboolean{astroph}
\setboolean{astroph}{true}

\shortauthors{Kov\'acs et al.}
\shorttitle{\hatcur\lowercase{b}}
\ifthenelse{\boolean{emulateapj}}{
    \newcommand{\titledag}{$\dagger$}
}{
    \newcommand{\titledag}{\dagger}
}

\begin{document}

\title{\hatcur\lowercase{b}: A 10.9-day Extrasolar Planet Transiting a 
Solar-type Star 
\altaffilmark{\titledag}}
\author{
	G.~Kov\'acs\altaffilmark{1},
	G.~\'A.~Bakos\altaffilmark{2,3},
	J.~D.~Hartman\altaffilmark{2},
	G.~Torres\altaffilmark{2},
	R.~W.~Noyes\altaffilmark{2},
	D.~W.~Latham\altaffilmark{2},
	A.~W.~Howard\altaffilmark{4},
	D.~A.~Fischer\altaffilmark{5},
	J.~A.~Johnson\altaffilmark{6},
	G.~W.~Marcy\altaffilmark{4},
	H.~Isaacson\altaffilmark{4}, 
	D.~D.~Sasselov\altaffilmark{2},
	R.~P.~Stefanik\altaffilmark{2},
	G.~A.~Esquerdo\altaffilmark{2},
    J.~M.~Fernandez\altaffilmark{2,8},
	B.~B\'eky\altaffilmark{2}, 
	J.~L\'az\'ar\altaffilmark{7},
	I.~Papp\altaffilmark{7},
	P.~S\'ari\altaffilmark{7}
}
\altaffiltext{1}{Konkoly Observatory, Budapest, Hungary}

\altaffiltext{2}{Harvard-Smithsonian Center for Astrophysics,
	Cambridge, MA, gbakos@cfa.harvard.edu}

\altaffiltext{3}{NSF Fellow}

\altaffiltext{4}{Department of Astronomy, University of California,
	Berkeley, CA}

\altaffiltext{5}{Department of Astronomy, Yale University, New Haven, CT
	06511}

\altaffiltext{6}{California Institute of Technology, Department of
	Astrophysics, MC 249-17, Pasadena, CA}



\altaffiltext{7}{Hungarian Astronomical Association, Budapest, 
	Hungary}

\altaffiltext{8}{Georg-August-Universit\"at G\"ottingen, Institut 
        f\"ur Astrophysik, G\"ottingen, Germany}

\altaffiltext{$\dagger$}{
	Based in part on observations obtained at the W.~M.~Keck
	Observatory, which is operated by the University of California and
	the California Institute of Technology. 
}


\begin{abstract}

We report the discovery of \hatcurb{}, a transiting extrasolar planet 
in the `period valley', a relatively sparsely-populated period regime 
of the known extrasolar planets. The host star, \hatcurCCgsc\,, is a 
\hatcurISOspec\, dwarf with $V=$\hatcurCCtassmvshort\,. It has a mass 
of \hatcurISOm\,\msun, radius of \hatcurISOr\,\rsun, effective 
temperature \hatcurSMEteff\,K, and metallicity $\feh =\hatcurSMEzfeh$. 
The planetary companion orbits the star with a period $P=\hatcurLCP\,d$, 
transit epoch $T_c =\hatcurLCT$ (BJD), and transit duration \hatcurLCdur\,d. 
It has a mass of \hatcurPPmlong\,\mjup, and radius of \hatcurPPrlong\,\rjup\ 
yielding a mean density of \hatcurPPrho\,\gcmc. At an age of 
\hatcurISOage\,Gyr, the planet is H/He-dominated and theoretical models 
require about $2$\% ($10$\mearth) worth of heavy elements to reproduce 
its measured radius. With an estimated equilibrium temperature of 
$\sim820$~K during transit, and $\sim1000$~K at occultation, \hatcurb{} 
is a potential candidate to study moderately cool planetary atmospheres 
by transmission and occultation spectroscopy. 
\end{abstract}



%
%
\section{Introduction}
\label{sec:introduction}

Ground-based transit surveys have been very successful at discovering 
short-period ($P<5$~days) transiting extrasolar planets (TEPs) 
since 2006. Because of the short duration of transit events (relative 
to the orbital period), candidates are more likely to be identified by 
higher duty cycle observations. The situation is especially demanding 
toward longer orbital periods, where the observational constraint 
couples with the considerably lower incidence rates of preferred orbital 
plane inclinations. The only way we can increase the probability of 
discovering long-period TEPs from the ground is to increase the duty 
cycle by establishing telescope networks spread in geographical 
longitude. Surveys such as WATTS/TrES and HATNet are designed with this 
goal in mind. For example, in the particular case of WATTS/TrES 
\citep{oetiker:2010}, for a typical observing season, the detection 
probability of a 10~day period TEP from a single observing site is 1--5\%. 
This increases to 10\% by using two sites of the network. It is clear 
that discovering long-period TEPs with ground-based surveys involves a 
large degree of serendipity even in the case of multisite observations.      

%
%
\begin{figure}[h]
\plotone{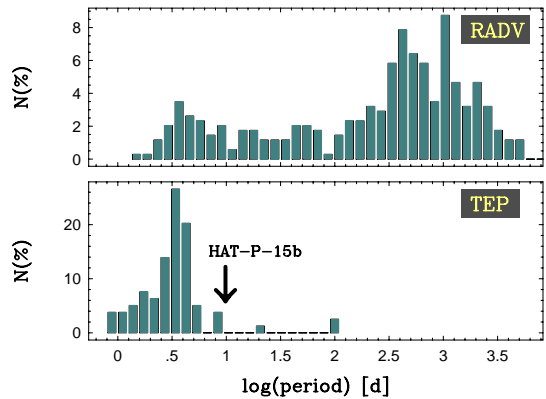}
\caption{
         Period distributions of the extrasolar planets discovered 
	 by radial velocity (RADV) method and by photometric transits 
	 (TEP -- a small fraction of TEPs was discovered by the 
	 radial velocity method and then later confirmed as TEPs). 
	 No TEPs are included in the distribution of the RADV planets. 
	 \hatcurb\, is situated at a relatively sparsely populated 
	 region in both diagrams. 	
\label{fig:pdistr}}
\end{figure}

In large part due to observational selection effects, from the more 
than $70$ TEPs\footnote{See http://exoplanet.eu/, as of May 10, 2010} 
known today, there are only 6 planets with orbital periods longer than 
6~days. Three of them 
\citep[CoRoT-4b, 6b and 9b, see][]{fridlund:2010, deeg:2010} were 
discovered by the CoRoT mission; two, HD~17156b \citep{barbieri:2007}, 
and HD~80606b \citetext{\citealp{moutou:2009}, \citealp{fossey:2009}, 
\citealp{garcia:2009} and \citealp{hebrard:2010}} came from the photometric 
followup observations of already known planets from radial velocity 
(RV) surveys. The only long-period planet originally discovered by a 
ground-based photometric survey is WASP-8b \citep{queloz:2010}, with 
a period of $8.16$~days. The period distribution of TEPs, displayed 
in the lower panel of \reffigl{pdistr}, shows the very sparse population 
of TEPs above $5$~days ($\log P > 0.7$). Interestingly, there are also 
relatively few radial velocity (RV) planets discovered in the period 
interval of $10$--$100$~days (top panel). The depletion of planets in 
this regime can be explained by the stellar-mass-dependent lifetime of 
the protoplanetary disk and its effect on migration 
\citetext{\citealp{burkert:2007} and \citealp{currie:2009}}. 

The 10.86~day period planet \hatcurb{} was discovered by the 
Hungarian-made Automated Telescope Network \citep[HATNet;][]{bakos:2004}. 
Operational since 2003, it has covered approximately 11\% of the 
Northern sky, searching for TEPs around bright stars 
($8\lesssim I \lesssim 12.5$\,mag). HATNet employs six wide field 
instruments: four at the Fred Lawrence Whipple Observatory (FLWO) 
in Arizona, and two on the roof of the Submillimeter Array hangar 
(SMA) of SAO in Hawaii.  Since 2006, HATNet has announced and 
published 14 TEPs. This work describes the fifteenth such discovery, 
the first TEP with a period above $10$~days discovered by a ground-based 
photometric survey. 

%
%
\section{Observations}
\label{sec:obs}
%
%
\subsection{Photometric detection}
\label{sec:detection}
The transits of \hatcurb{} were detected with the HAT-6 telescope in
Arizona and the HAT-9 telescope in Hawaii. The region around
\hatcurCCgsc{}, a field internally labeled as G\hatcurfield, was
observed on a nightly basis between 2005 September and 2006 February,
whenever weather conditions permitted. We gathered 8538 images with an 
exposure time of 5 minutes and with readout time and other operations 
yielding a 5.5 minute cadence. Each image contained approximately 
55,000 stars down to $I\sim14$~mag. For the brightest stars in the 
field, we achieved a per-image photometric precision of 2.5\,mmag.

The calibration of the HATNet frames was performed by utilizing 
standard procedures as described in our earlier papers. In brief, 
star detection, astrometry and aperture photometry were performed on 
the calibrated images by using the Two Micron All Sky Survey (2MASS) 
catalog \citep{skrutskie:2006} and the astrometric routine of 
\cite{pal:2006}. The resulting \lcs\ were decorrelated from 
systematics using the External Parameter Decorrelation technique 
\citep[EPD, ][]{bakos:2010} and the Trend Filtering Algorithm 
\citep[TFA, ][]{kovacs:2005}. The \lcs{} were searched for periodic 
box-like signals using the Box Least-Squares method 
\citep[BLS, ][]{kovacs:2002}. 

\begin{figure}[!ht]
\plotone{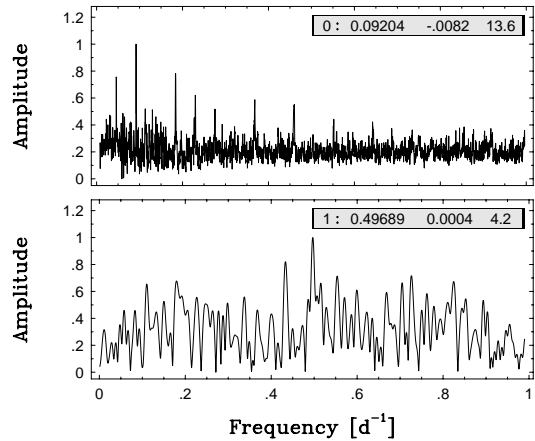}
\caption{
        {\em Upper panel:} BLS frequency spectrum of the TFA-filtered 
        HATNet data of \hatcur{}. {\em Lower panel:} DFT spectrum of 
        the data prewhitened by the best-fit trapezoidal signal derived 
        from the BLS analysis above. Both spectra are normalized to 
        1.0 at the highest peak. Insets in the panels show, from 
        left to right: prewhitening order, peak frequency, transit 
        depth (BLS) or Fourier amplitude (DFT) and signal-to-noise 
        ratio. 
\label{fig:hatnet-spec}}
\end{figure}

We detected a significant signal in the \lc{} of \hatcurCCgsc{} 
(\hatcur, also known as \hatcurCCtwomass{}; 
$\alpha =$\hatcurCCra\,, $\delta =$\hatcurCCdec\,; J2000; 
$V=$\hatcurCCtassmvshort\,; \citealp{droege:2006}). As shown in 
the upper panel of \reffigl{hatnet-spec}, the detection of \hatcur\, 
is clear, in spite of its long period. The occurrence of subharmonic and 
harmonic components is an artefact of BLS (or any other phase-folding 
period search algorithms with arbitrary signal shape), and indicates 
the significance of the detection. During the total time span of 135~days, 
we had three transit events with nearly full, and two more with very 
limited transit coverages, comprising altogether some 130 data points 
within the transit. Except for one marginally overlapping event, all 
of them were separately caught by HAT-6 and HAT-9, stressing the 
importance of using a network in the search for long-period TEPs. 

We searched for additional components in the signal by subtracting 
the best-fitting trapezoidal signal corresponding to the peak frequency 
of the above BLS analysis. We did not find any component (transit 
[BLS] or Fourier [DFT]) that appeared significant. The null 
result for the DFT search is shown in the lower panel of 
\reffigl{hatnet-spec}. In agreement with the low chromospheric activity 
suggested by the spectroscopic data (see \refsecl{hispec}) we can 
exclude any coherent Fourier signal above $\sim5$~mmag.  

%
%

%
%
\begin{figure}[!ht]
\plotone{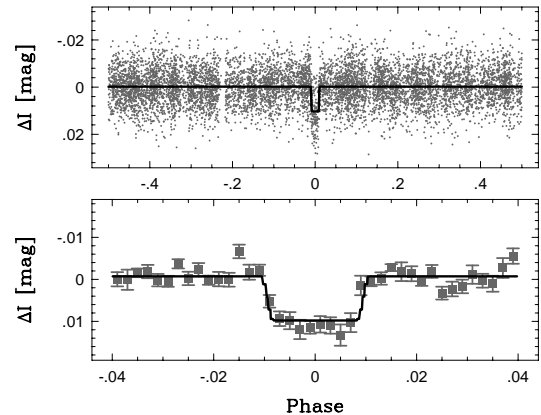}
\caption{
	{\em Upper panel:} Unbinned \lc{} of \hatcur{} including all 
        8538 instrumental \band{I} $5.5$ minute cadence measurements 
	obtained with the HAT-6 and HAT-9 telescopes of HATNet, 
	and folded with the period of $P=$\hatcurLCPshort\,days. 
	{\em Lower panel:} Folded/binned light curve, zoomed to the 
        transit. We used 500 bins in the full period. Error bars show 
        the $1\sigma$ errors of the bin averages. The solid line shows 
	the transit model fit in the simplified \cite{mandel:2002} 
	approximation \citep[without limb darkening, see][]{bakos:2010}  
	as derived \refsecl{globmod}.
\label{fig:hatnet}}
\end{figure}

\reffigl{hatnet} shows the folded HATNet \lc{}, on which is superposed 
the model fit to be discussed in \refsecl{globmod}. The flat out-of-transit 
part of the \lc{} confirms the null result on the presence of additional 
components, in particular of those with periods commensurable with the 
orbital periods 
\citep[secondary eclipse or ellipsoidal variation, see][]{sirko:2003}, 
which might suggest a blended eclipsing binary. The derived transit 
parameters of the HATNet \lc{} are also in good agreement with the 
considerably more accurate followup data to be presented in 
\refsecl{phot}. We obtained a period of 
$P=10.8645\pm0.0046$~days, a moment of transit center of 
$T_c=2453649.974\pm0.037$~[HJD], a relative transit duration (first 
to last contact divided by the period) of $q=0.0188\pm0.0018$ and a 
transit depth of $0.0110\pm0.0008$~mag. The HATNet \lc{} also predicts 
a short ingress time of $(0.063\pm0.065)Pq$~days, again consistent 
with the followup data. In spite of the relatively large $1\sigma$ 
statistical errors quoted, by using the HATNet ephemeris we get a 
difference of only $0.08$~days between the transit time predicted 
from the HATNet light curve and that observed from the followup data 
(\refsecl{phot} and Table~\ref{tab:planetparam}).

%
%
\subsection{Reconnaissance Spectroscopy}
\label{sec:recspec}
%

The first step in our follow-up investigation of planet host candidates 
is usually to gather low-S/N, high-resolution spectra to obtain a basic 
characterization of the host star and exclude obvious blend configurations. 
As in most of our earlier works, we used the CfA Digital Speedometer 
\citep[DS;][]{latham:1992}, mounted on the \flwos\ telescope. For a 
description of DS for reconnaissance spectroscopy we refer to 
\cite{latham:2009}.  

We obtained \hatcurDSnumspec\, DS spectra of \hatcur{} in 2006 spanning 
\hatcurDSspan\,days. The RV measurements showed an RMS residual of 
\hatcurDSrvrms\,\kms\,. Considering that the overall formal error of 
the individual measurements is the same as the RMS, the observations 
suggest that there is no detectable RV variation above the \kms\, level. 
This rules out a short period stellar companion. Furthermore, the 
spectra were single-lined, showing no evidence of more than one star 
in the system. Atmospheric parameters for the star, including the 
effective temperature $\teffstar = \hatcurDSteff\,K$, surface gravity 
$\loggstar (\rm CGS) = \hatcurDSlogg$, and projected rotational velocity 
$\vsini= \hatcurDSvsini\,\kms$, were derived as described by 
\cite{torres:2002}. The effective temperature and surface gravity 
correspond to a late G dwarf. The mean heliocentric RV of \hatcur\ is 
\hatcurDSgamma\,\kms. These stellar parameters, the absence of large 
velocity dispersion and the lack of any stellar contamination indicated 
that the target warranted further follow-up studies. 

%
%
\subsection{High resolution, high S/N spectroscopy}
\label{sec:hispec}

To obtain high-resolution and high-S/N spectra for characterizing 
the RV variations and to determine the stellar parameters with higher 
precision, we used the HIRES instrument \citep{vogt:1994} on the Keck~I 
telescope located on Mauna Kea, Hawaii. We obtained $24$ exposures with 
an iodine cell and one iodine-free template. The observations were made 
on 21 nights during a number of observing runs between 2007 August 24 
and 2009 December 28.

%
%
\begin{figure}[ht]
\ifthenelse{\boolean{emulateapj}}{
	\plotone{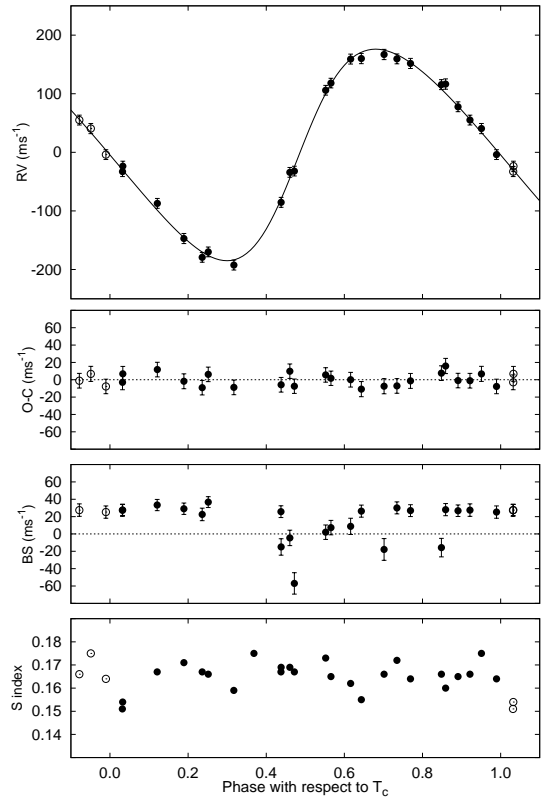}
}{
	\includegraphics[scale=0.75]{\hatcurhtr-rv.eps}
}
\caption{
    {\em Top:} RV measurements from Keck for \hatcur{}, shown as a function 
    of orbital phase, using our best-fit model (see \refsecl{analysis}). 
    Zero phase is defined at the transit midpoint. The center-of-mass 
    velocity has been subtracted. Note that the error bars include the 
    stellar jitter of \hatcurRVjitter\,\ms, added in quadrature to the 
    formal errors given by the spectrum reduction pipeline. 
    {\em Second panel from the top:} phased residuals after subtracting 
    the orbital fit. The rms of the residuals is \hatcurRVrms\,\ms.
    {\em Third panel from the top:} phased bisector spans (BS, mean 
    value subtracted). The low-lying scattered points are attributed  
    to the bias caused by the Moon (see \refsecl{blend}). Please note 
    that the three most extreme outliers have been omitted, in order 
    to show sufficient amount of detail.  
    {\em Bottom:} Calibrated S activity index values computed from 
    the Keck spectra. 
    Note the different vertical scales of the panels. Open circles are 
    for the re-plotted values from the [0,1] phase interval. 
	\label{fig:rvbis}
}
\end{figure}

%
\ifthenelse{\boolean{emulateapj}}{
    \begin{deluxetable*}{lrrrrrr}
}{
    \begin{deluxetable}{lrrrrrr}
	\tabletypesize{\scriptsize}
}
\tablewidth{0pc}
\tablecaption{
	Relative radial velocity, bisector and activity indices of \hatcur{}.
	\label{tab:rvs}
}
\tablehead{
	\colhead{BJD} & 
	\colhead{RV\tablenotemark{a}} & 
	\colhead{\ensuremath{\sigma_{\rm RV}}\tablenotemark{b}} & 
	\colhead{BS} & 
	\colhead{\ensuremath{\sigma_{\rm BS}}} & 
	\colhead{S\tablenotemark{c}} & 
	\colhead{\ensuremath{\logrhk}\tablenotemark{c}}\\
	\colhead{\hbox{~~~~(2,454,000$+$)~~~~}} & 
	\colhead{(\ms)} & 
	\colhead{(\ms)} &
	\colhead{(\ms)} &
    \colhead{(\ms)} &
	\colhead{} &
	\colhead{}
}
\startdata
$ 337.11515 $ \dotfill & $  -169.94 $ & $     2.28 $ & $    36.70 $ & $     6.25 $ & $    0.1660 $ & $ -4.941 $\\
$ 339.13623 $ \dotfill & \nodata      & \nodata      & $    25.72 $ & $     6.78 $ & $    0.1670 $ & $ -4.935 $\\
$ 339.14209 $ \dotfill & $   -85.37 $ & $     2.75 $ & $   -14.98 $ & $     9.46 $ & $    0.1690 $ & $ -4.923 $\\
$ 344.05984 $ \dotfill & $    77.73 $ & $     2.40 $ & $    26.77 $ & $     6.66 $ & $    0.1650 $ & $ -4.947 $\\
$ 345.13121 $ \dotfill & $    -3.91 $ & $     2.35 $ & $    25.23 $ & $     7.10 $ & $    0.1640 $ & $ -4.953 $\\
$ 370.97691 $ \dotfill & $  -234.64 $ & $     2.76 $ & $  -114.84 $ & $    15.74 $ & $    0.1750 $ & $ -4.890 $\\
$ 371.97469 $ \dotfill & $   -34.16 $ & $     2.29 $ & $    -4.51 $ & $     8.90 $ & $    0.1690 $ & $ -4.923 $\\
$ 372.10061 $ \dotfill & $   -32.06 $ & $     2.10 $ & $   -57.00 $ & $    12.36 $ & $    0.1670 $ & $ -4.935 $\\
$ 372.97042 $ \dotfill & $   105.93 $ & $     2.15 $ & $     2.03 $ & $     8.37 $ & $    0.1730 $ & $ -4.901 $\\
$ 373.11653 $ \dotfill & $   118.15 $ & $     2.03 $ & $     7.31 $ & $     8.35 $ & $    0.1650 $ & $ -4.947 $\\
$ 397.91546 $ \dotfill & $   115.53 $ & $     2.75 $ & $   -15.73 $ & $    10.65 $ & $    0.1660 $ & $ -4.941 $\\
$ 399.03020 $ \dotfill & $    40.57 $ & $     3.37 $ & $  -102.32 $ & $    19.47 $ & $    0.1750 $ & $ -4.890 $\\
$ 427.98281 $ \dotfill & $   159.16 $ & $     2.68 $ & $     8.79 $ & $     9.32 $ & $    0.1620 $ & $ -4.965 $\\
$ 428.91259 $ \dotfill & $   167.01 $ & $     3.46 $ & $   -17.89 $ & $    12.52 $ & $    0.1660 $ & $ -4.941 $\\
$ 455.93552 $ \dotfill & $  -147.17 $ & $     3.07 $ & $    29.07 $ & $     6.58 $ & $    0.1710 $ & $ -4.912 $\\
$ 460.87067 $ \dotfill & $   160.04 $ & $     3.45 $ & $    26.32 $ & $     6.95 $ & $    0.1550 $ & $ -5.012 $\\
$ 548.76912 $ \dotfill & $   159.50 $ & $     2.44 $ & $    30.05 $ & $     6.95 $ & $    0.1720 $ & $ -4.906 $\\
$ 728.03347 $ \dotfill & $  -179.29 $ & $     2.33 $ & $    22.53 $ & $     7.25 $ & $    0.1670 $ & $ -4.935 $\\
$ 778.93314 $ \dotfill & $    55.05 $ & $     2.66 $ & $    27.48 $ & $     7.17 $ & $    0.1660 $ & $ -4.941 $\\
$ 809.86948 $ \dotfill & $   151.70 $ & $     3.05 $ & $    26.90 $ & $     6.85 $ & $    0.1640 $ & $ -4.953 $\\
$ 810.84709 $ \dotfill & $   116.50 $ & $     3.53 $ & $    28.02 $ & $     7.00 $ & $    0.1600 $ & $ -4.978 $\\
$ 1109.13606$ \dotfill & $  -192.39 $ & $     2.72 $ & $   -83.90 $ & $    14.61 $ & $    0.1590 $ & $ -4.985 $\\
$ 1192.94775$ \dotfill & $   -32.84 $ & $     2.99 $ & $    27.35 $ & $     6.82 $ & $    0.1510 $ & $ -5.041 $\\
$ 1192.95593$ \dotfill & $   -23.64 $ & $     2.92 $ & $    27.56 $ & $     6.57 $ & $    0.1540 $ & $ -5.019 $\\
$ 1193.91529$ \dotfill & $   -87.16 $ & $     2.42 $ & $    33.33 $ & $     6.45 $ & $    0.1670 $ & $ -4.935 $\\
\enddata
\tablenotetext{a}{
	The fitted zero-point that is on an arbitrary scale (denoted as
	$\gamma_{\rm rel}$ in \refsecl{globmod}) has been subtracted 
	from the velocities.
}
\tablenotetext{b}{
        The values for \ensuremath{\sigma_{\rm RV}} do not include
        the jitter.
}
\tablenotetext{c}{
        Activity indices calibrated to the scale of \citet{duncan:1991}. 
}
\ifthenelse{\boolean{rvtablelong}}{
	\tablecomments{
        The exposure on BJD~2,454,339.13623 was iodine-free, 
	therefore, no RV item is given. On BJD~2,454,370.97691 
	the scattered moonlight affected the derived parameters 
        very strongly, including RV. As a result, we left out 
        this observation from the analysis.   
	}
}{
    \tablecomments{
		Note that for the iodine-free template exposures we do not
		measure the RV but do measure the BS and S index. Such template
		exposures can be distinguished by the missing RV value. 
        This table is presented in its entirety in the electronic edition
        of the Astrophysical Journal. A portion is shown here for guidance
        regarding its form and content.
	}
} 
\ifthenelse{\boolean{emulateapj}}{
    \end{deluxetable*}
}{
    \end{deluxetable}
}

The instrumental setup and reduction method used for HIRES are the 
same as given in our earlier papers and described in more detail 
by \cite{marcy:1992} and \cite{butler:1996}. The final RV data and
their errors are listed in \reftabl{rvs}. The folded data are plotted 
in \reffigl{rvbis}, with our best fit (see \refsecl{analysis}) 
superimposed. Not plotted is the data point taken on BJD~$2454370.9769$, 
due to its outlier status (most probably due to the large contamination 
from the Moon -- see \refsecl{blend}). 

For each spectrum we measured the S and \rhk\, chromospheric
activity indices from the emission in the CaII H and K line cores. 
The S index is computed following the prescription given by 
\cite{duncan:1991}, after matching each spectrum to a reference 
spectrum using a transformation that includes a wavelength shift 
and a flux scaling that is a polynomial as a function of wavelength. 
The transformation is determined on regions of the spectra that 
are not used in computing the indices. The \rhk\, index, as 
described by \cite{noyes:1984}, uses the S-value and the $(B-V)$ 
color of the star to determine the amount of flux in the H and K 
line cores as a fraction of the total energy output of the star. 
The use of $(B-V)$ color in defining \rhk\, allows for comparison 
of chromospheric activity across spectral types. The Mt. Wilson 
survey sampled mostly bright, Sun-like stars, therefore \rhk\, 
is only valid for the color range $0.4\la B-V\la 1.2$. Both S 
and \rhk\, have been calibrated to the Mt. Wilson scale as given 
by \citet{duncan:1991}. 

The RMS of the residuals of the model fit to the radial velocities 
described later is \hatcurRVrms\,\ms, which is larger than the 
expected scatter of $\sim 2.8$\,\ms, based on the formal errors. 
Assuming that the difference is due to spot or other chromospheric 
activity of the star and that the errors add quadratically, we get 
a value of \hatcurRVjitter\,\ms\ for the RMS of the ``stellar jitter''. 
This is more than an order of magnitude larger than the expected 
jitter of $0.38$\,\ms, for a star similar to the Sun 
(see \citealp{makarov:2009}).\footnote{However, we note that the 
average of the measured values of the stellar jitter in quiet 
solar-type stars is about $2$\ms\, \citep{wright:2005}. Nevertheless, 
this is still a few times lower than the jitter of \hatcur{}.} 
Indeed, the parameters of \hatcur{}\, (see \reftabl{stellar}) are 
very close to those of the Sun, albeit its metallicity is slightly 
higher and its estimated age is larger, too. The activity indices 
also imply that \hatcur{} is a very quiet star. The average and 
standard deviation of S and \logrhk\, are $(0.165,0.006)$ and 
$(-4.95,0.04)$, respectively, matching very nicely the solar values 
\citep{hall:2009}. Furthermore, the low chromospheric activity is 
also in agreement with the sub-millimagnitude upper limit obtained 
for the photometric variability (\refsecl{detection}). This is what 
we expect for a quiet solar-type star \citep{makarov:2009}. All in all, 
the extra RV scatter is most probably not related to the `traditional' 
source of stellar jitter, but may come from other sources, namely: 
(i) observational errors due to instrumental or environmental effects; 
(ii) light contamination from the Moon; 
(iii) perturbation due to other bodies in the system. 
Although the contribution to the error budget from (i) may be larger 
than estimated, here we assume that this is not very likely (it would 
imply a factor $2.5$ difference between the formal and the true errors). 
Possibility (ii) has been examined by omitting the data points that are 
most probably affected by scattered moonlight -- see \refsecl{blend} for 
demonstrating the strong effect of moonlight on the observed distortion 
of the bisector span (BS). This test resulted in a decrease in the RMS 
of up to 30\%, but only at the price of omitting a large number (up to 11) 
of RV observations that have associated BS values affected by the Moon, 
but otherwise fit the single-planet Keplerian orbit fairly well.
All of this indicates the possible presence of one or more additional 
planets in the system. However, an extensive search for the signature 
of an additional planet in the RV data did not yield a significant 
detection. 
 
%
%
\subsection{Photometric follow-up observations}
\label{sec:phot}
%

%
%
\begin{figure}[!ht]
\includegraphics[scale=0.75]{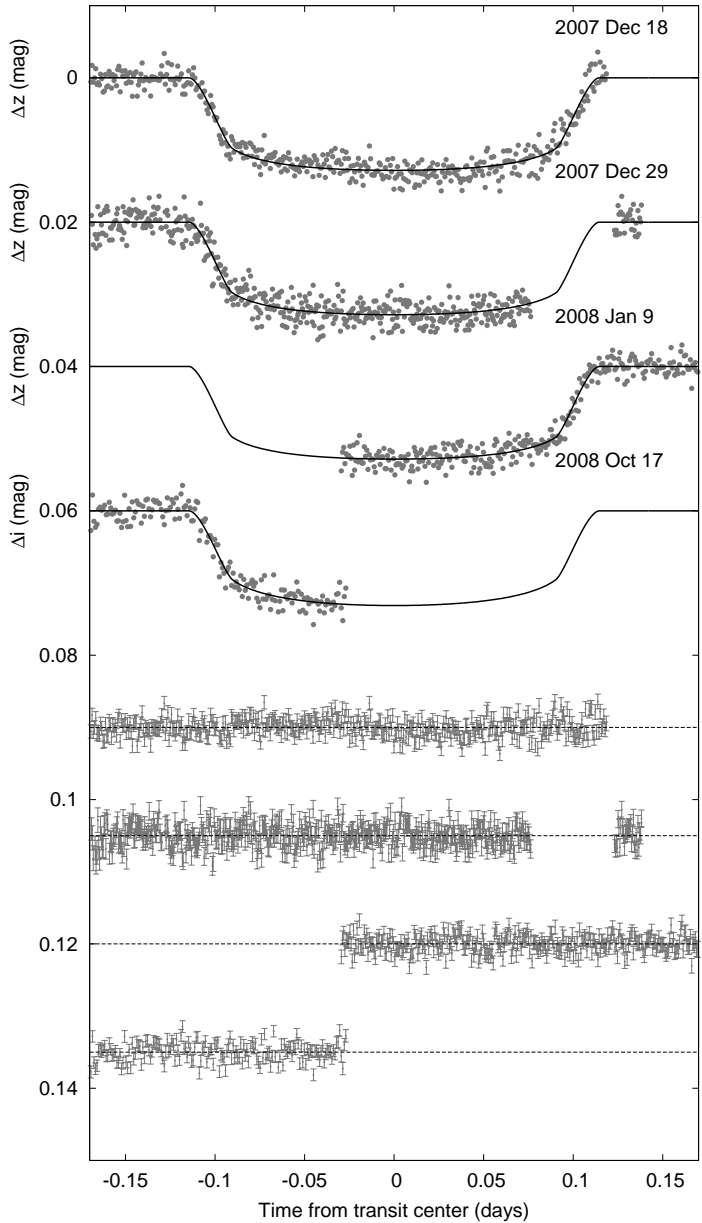}
\caption{
	Unbinned instrumental $z$ and $i$ band transit \lcs{},
        acquired with KeplerCam at the \flwof{} telescope.
        At the bottom of the figure we show the residuals from the 
	fit. Error bars represent the photon and background shot 
	noise, plus the readout noise. 
\label{fig:lc}}
\end{figure}

To confirm the transit signal and obtain high-precision light curves
for modeling the system, we conducted photometric follow-up observations 
with KeplerCam \citep{szentgyorgyi:2005} at the \flwof{} telescope. 
Our journal of observations is given in \reftabl{phfu-journal}.  

%
%
\begin{deluxetable}{lcccc}
\tablewidth{0pc}
\tablecaption{Journal of photometric follow-up observations of \hatcur{} 
\label{tab:phfu-journal}}
\tablehead{
	\colhead{Date} & 
	\colhead{\ensuremath{N_{\rm frames}}} & 
	\colhead{Cad. [sec]} & 
	\colhead{Exp. [sec]} &
	\colhead{Filter}}
\startdata
2007 Dec. 18 & $608$ & $60$ & $45$ & ${\rm Sloan}~z$ \\
2007 Dec. 29 & $514$ & $45$ & $30$ & ${\rm Sloan}~z$ \\
2008 Jan. 09 & $400$ & $60$ & $45$ & ${\rm Sloan}~z$ \\
2008 Oct. 17 & $276$ & $75$ & $60$ & ${\rm Sloan}~i$ \\ [-1.5ex]
\enddata
\end{deluxetable}

%

%
%
\begin{figure}[!ht]
\plotone{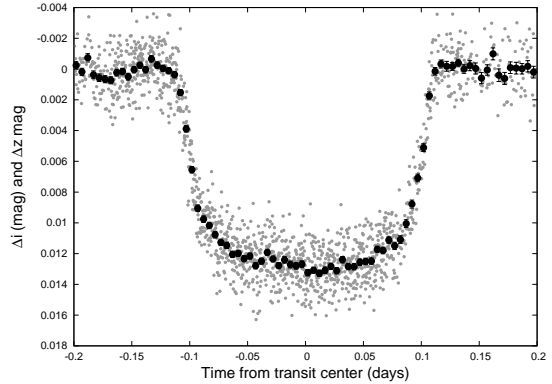}
\caption{
	Folded/binned Sloan \band{z,i} \lc{} of \hatcur{}. 
	The data have been obtained by KeplerCam at the \flwof{} 
	telescope. The barely visible error bars are $1\sigma$ 
	errors of the bin averages. 
\label{fig:fold-flwo}}
\end{figure}

The reduction of the images was performed as follows. After bias and 
flat-field calibration, we derived a second order astrometric 
transformation between the $\sim 400$ brightest stars and the 2MASS 
catalog, as described in \citet{pal:2006}, yielding a residual of 
$\la 0.2$ pixels.
Aperture photometry was then performed on the resulting fixed positions, 
using a range of apertures. The instrumental magnitude transformation 
was done in two steps: first, all magnitude values were transformed to 
the photometric reference frame (selected to be the sharpest image), 
using the individual Poisson noise error estimates as weights. Next, the
magnitude fit was repeated using the mean individual \lc{} magnitudes
as reference and the RMS of these \lcs{} as weights. In both of these
magnitude transformations we excluded from the fit the target star
itself and the $3\sigma$ outliers. 

We performed EPD and TFA to remove trends simultaneously with the 
light curve modeling 
\citep[for more details, see \refsecl{analysis} and][]{bakos:2010}. 
In conducting this analysis we reject $3\sigma$ outliers from the 
best-fit transit \lc{} models. For the 2007 December 18, 2007 December 29, 
2008 January 9, and 2008 October 17 observations we rejected 4, 4, 1 
and 1 images respectively. 

Of the several apertures used each night, we chose the one yielding the
smallest fit RMS for the \lc. The final \lcs{} are shown in the upper
plots of \reffigl{lc}. To exhibit the resulting photometric accuracy, 
when all data are used, we plot the folded $z$ and $i$ observations in 
\reffigl{fold-flwo}. Although we plot light curves obtained in two 
different wavebands, there is only a negligible difference between the 
light curves in these two bands due to the wavelength dependence of the 
stellar limb darkening. The overall standard deviation of the bin averages 
is $\sim0.2$~mmag per bin length of $\sim7$~min. 

%
%
\begin{deluxetable}{lrrrr}
\tablewidth{0pc}
\tablecaption{Photometric follow-up of \hatcur\label{tab:phfu}}
\tablehead{
	\colhead{BJD} & 
	\colhead{Mag\tablenotemark{a}} & 
	\colhead{\ensuremath{\sigma_{\rm Mag}}} &
	\colhead{Mag(orig)\tablenotemark{b}} & 
	\colhead{Filter} \\
	\colhead{\hbox{~~~~(2,400,000$+$)~~~~}} & 
	\colhead{} & 
	\colhead{} &
	\colhead{} & 
	\colhead{}
}
\startdata
$ 54453.57510 $ & $   0.00189 $ & $   0.00122 $ & $  10.64060 $ & $ z$\\
$ 54453.57987 $ & $   0.00257 $ & $   0.00121 $ & $  10.64190 $ & $ z$\\
$ 54453.58575 $ & $  -0.00044 $ & $   0.00098 $ & $  10.63850 $ & $ z$\\
$ 54453.58644 $ & $  -0.00011 $ & $   0.00098 $ & $  10.63880 $ & $ z$\\
$ 54453.58713 $ & $   0.00108 $ & $   0.00098 $ & $  10.64010 $ & $ z$\\
[-1.5ex]
\enddata
\tablenotetext{a}{
	The out-of-transit level has been subtracted. These magnitudes 
	have resulted from the EPD and TFA procedures, carried out
	simultaneously with the transit fit.
}
\tablenotetext{b}{
	These are raw magnitude values without the application of 
	the EPD and TFA procedures.
}
\tablecomments{
    This table is available in a machine-readable form in the
    online journal. A portion is shown here for guidance regarding
    its form and content.
}
\end{deluxetable}

%
%
\section{Excluding blend scenarios}
\label{sec:blend}

Here we examine if the distortion of the spectral line correlation 
profiles (characterized by the corresponding bisector spans, BS) 
show any variation correlated with the orbital period, radial velocity 
(or their model-fit residuals) and stellar activity indices. It is 
well known that these tests are sensitive to distortions caused by 
hidden background (or physically related) unresolved eclipsing binaries 
or stellar activity \citep{queloz:2001, torres:2007}. 

\reffigl{rvbis} shows that the BS distribution as a function of 
orbital phase is nearly flat, except for several low points whose 
deviations are apparently not correlated either with the RV residuals 
or with the activity index. Following our earlier work related to 
the verification of HAT-P-12b \citep{hartman:2009} we investigated 
the contamination caused by the moonlight. By adopting the same 
definition of the sky contamination factor SCF as in 
\cite{hartman:2009}, we found a suggestive correlation between SCF 
and BS, but with several outliers. We attribute this to the fact 
that the simple formula given in \cite{hartman:2009} only holds 
when the RV difference between the Moon and the star is greater 
than the width of the cross-correlation function (CCF) profile. 
When the RV difference is close to zero, the component of the CCF 
due to scattered moonlight will affect both sides of the CCF component 
due to the star, reducing the effect of sky contamination on the 
BS. We addressed this by direct modeling of the measured BS. 

%
%
\begin{figure} [ht]
\plotone{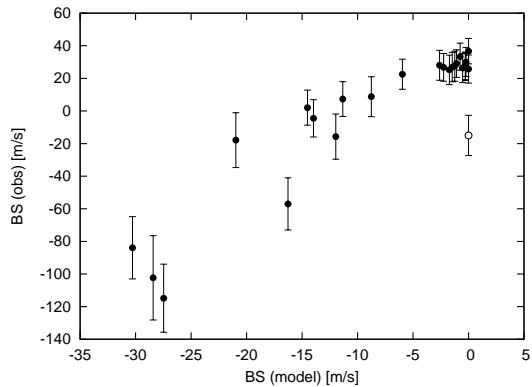}
\caption{
         Effect of the Moon on the variation of the bisector span BS. 
	 The model BS values have been derived by synthesizing the 
	 simplified correlation spectra of the star and that of the 
	 Moon. Depending on the Moon phase/intensity, target-Moon 
	 separation and relative moon velocity we get various degrees 
	 of contamination, showing significant correlation with the 
         observed BS variation. The apparent outlier corresponding 
         to the measurement taken on BJD~2454339.14209 is shown by 
         an open circle.   
\label{fig:bs-moon}}
\end{figure}

For each spectrum we generated an expected CCF, taken to be the sum
of two Lorentzian functions, one for the target and the other for 
the scattered moonlight. The radial velocity separation of the two
Lorentzians, and their relative intensities, were determined as in
\cite{hartman:2009}. These CCFs were used to compute BS(model) in 
the same way as it is done for the observed CCFs. The result is 
shown in \reffigl{bs-moon}. With the possible exception of the 
measurement taken on BJD~2454339.14209, the real BS seems to 
correlate well with the simulated BS, confirming that the former 
is affected by moonlight to varying degrees. The cause of the one 
outlier mentioned above is not known, since during the time of the 
observation the Moon was below the Horizon. 

It is also seen that the expected moon-contaminated BS variations 
are a factor of a few lower than the observed BS variations. 
Presumably this is due to underestimation of the height of the 
sky contamination peak and/or the use of an incorrect Lorentzian 
width. Clouds will generally increase the sky background beyond the 
estimations, and airglow (which is completely ignored here) will 
also add emission lines to the spectrum at RV$=0.0$\,\ms. In the 
BS analysis we try to clip obvious airglow lines when cleaning 
cosmic rays, but this is by no means perfect. 

Although visual inspection of \reffigl{rvbis} suggests that there 
are no correlations between RV, BS and S, we examined this problem 
more rigorously by computing false alarm probabilities (FAPs) for 
various combinations, including data selection based on moonlight 
contamination. Most FAPs are above $30$--$50$\% with associated 
correlation coefficients CC less than $0.2$ (in absolute value). 
The least uncorrelated sets are the RV residuals and the BS values, 
obtained by the omission of the data points affected by the Moon 
(see the $11$ points in \reffigl{bs-moon} with 
${\rm BS(model)} < -5$\ms\,). For the remaining $13$ BS, RV(residual) 
data points we obtained a FAP of $3$\% and CC of $+0.6$. Considering 
the low number of data points in this test, and that the correlation 
heavily relies on a single point, these values are still regarded 
as characteristics of a random correlation. It is worth pointing 
out that the relative dispersion of the BS values is fairly small 
if we omit the BS values contaminated by the Moon. The ratio of the 
standard deviation of the RV data (without subtracting the orbital 
motion) and that of the BS points is only $0.03$. We conclude that 
all evidence strongly support the conclusion that the source of the 
observed RV variation is the orbital motion of a planetary companion 
around \hatcur{}. 

%
%
\section{Analysis}
\label{sec:analysis}

For the determination of the physical parameters of the planet, 
first we have to derive those of the host star. Since the stellar 
density is tied to the transit and orbital parameters, we use an 
iterative process in which we utilize several observational and 
theoretical model components, such as: the high-resolution template 
spectrum obtained with Keck/HIRES; stellar evolution and atmosphere 
models; global/simultaneous analysis of the \lc{} and RV data 
to derive orbital and relative planet parameters. We end up with 
absolute stellar and planet parameters and corresponding errors 
estimated from the Monte Carlo simulations performed within the 
iteration process.  

%
%
\subsection{Properties of the parent star}
\label{sec:stelparam}

We derived the stellar atmosphere parameters from the Keck/HIRES 
template spectrum as given in our earlier papers 
\citep[e.g.,][]{torres:2010} and described in more detail in 
\cite{valenti:1996} and \cite{valenti:2005}. This yielded the 
following {\em initial} values and uncertainties:\footnote{Please 
note that the errors are formal and that we have conservatively 
increased the error of \teff\, and \feh\, by a factor of two.}
effective temperature $\teffstar=\hatcurSMEiteff$\,K, 
stellar surface gravity $\loggstar(cgs)=\hatcurSMEilogg$\,, 
metallicity $\feh=\hatcurSMEizfeh$\,dex, and 
projected rotational velocity $\vsini=\hatcurSMEivsin\,\kms$. 
The differences compared to our previous DS estimates 
(\refsecl{recspec}) can be attributed in part to fixing the 
metallicity (to that of the Sun) in the DS analysis. 

Following our earlier practice, at this stage we can utilize the 
fact that in the case of a low-mass companion, the directly measured 
transit parameters (most importantly \arstar, the ratio of the 
semi-major axis to the stellar radius) tightly constrain \rhostar, 
the mean stellar density and thereby also \loggstar, entering in 
the stellar atmosphere models \citep{seager:2003, sozzetti:2007}. 

Considering the coupling between the orbital and stellar parameters, 
in our standard iterative procedure we first adopt the values of 
\teffstar\,, \feh\,, and \loggstar\, from the initial spectral 
analysis to fix the quadratic limb-darkening coefficients as given 
by \cite{claret:2004} (these coefficients are needed to model the 
\lcs\, in the observed {\it i} and {\it z} bands). The \lc\, modeling 
yields \arstar\, (see \refsecl{globmod}). We then use \arstar\,, 
\teffstar\, and \feh\, to estimate \mstar\ and \rstar\ by comparison 
with the stellar evolution models of \citet{yi:2001}. To obtain 
proper error estimates, the procedure was repeated a large number 
of times with \teffstar\, and \feh, drawn from Gaussian distributions, 
whereas \arstar\, from the one derived in the global analysis of 
\refsecl{globmod}. Although the resulting surface gravity of 
$\loggstar =\hatcurISOlogg$ is only slightly different from that 
derived in the initial spectral analysis, to see the effect of this 
change we performed a second iteration by holding \loggstar\ fixed 
at the value above in the new spectral analysis, and adjusting only 
\teffstar, \feh, and \vsini. This gave
$\teffstar = \hatcurSMEiiteff$\,K, 
$\feh = \hatcurSMEiizfeh$, and 
$\vsini = \hatcurSMEiivsin$\,\kms, 
which we adopt as final atmospheric values for the star. The additional 
stellar parameters have been derived from the evolution models 
(see \reftabl{stellar}).

\begin{figure}[!ht]
\plotone{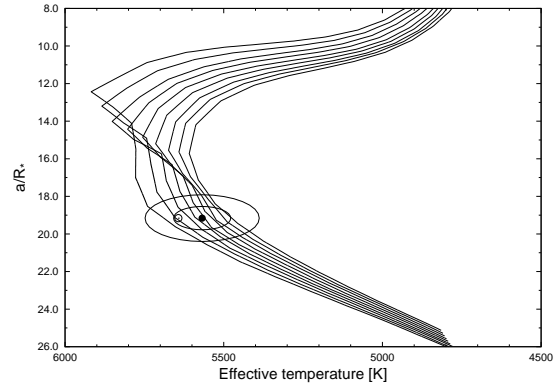}
\caption{
	Stellar isochrones of \citet{yi:2001} in the 
	\teff\,---$(\arstar$) plane for metallicity
        \feh=\hatcurSMEiizfehshort\ and ages of 5.0, 5.5, 6.0, 6.5, 
	7.0, 7.5, 8.0, 8.5 and 9.0\,Gyr, from left to right. 
	The initial value of \teffstar\ and \arstar\ is marked by 
	an open circle, whereas the final choice is shown as a 
	solid dot, encircled by the $1\sigma$ and $2\sigma$ confidence 
	ellipsoids.
\label{fig:iso}}
\end{figure}

In \reffigl{iso} we plot the model isochrones of \citet{yi:2001} 
for \feh=\hatcurSMEzfehshort\, with the final choice of $\teffstar$ 
and \arstar\, marked, and encircled by the 1$\sigma$ and 2$\sigma$ 
confidence ellipsoids.

In supplementing the above stellar parameters, we note that the 
average values of the stellar activity indices (see \reftabl{rvs}) 
also allow us to get independent estimates on the stellar age and 
rotation period. Since \hatcur{} is very similar to the Sun, it is 
expected that both of these parameters are very close to those of 
the Sun. Indeed, from Fig.~6 of \cite{mamajek:2008}, with 
\logrhk$=-4.95$ we get an age interval of $4.0$--$8.0$\,Gyr, in good 
agreement with our estimate based on stellar evolution isochrones 
(see \reftabl{stellar}). Similarly, the expected rotation rate of 
$\sim 30$~days (based on the \logrhk\, calibration of the Rossby 
number and an assumed overall convective turnover time of 
$\sim15$\,days -- see \citealp{noyes:1984}, \citealp{mamajek:2008}) 
is in good agreement with the value of $\sim 27$~days derived 
from our stellar parameters and spectroscopic $\vsini$ of 
$\hatcurSMEiivsin\,\kms$. We note, however, that there is no sign 
of significant power of any periodicity in the frequency spectra 
of S and \logrhk\,. 

%
%

As discussed below, \hatcur{} is projected against the edge of 
the Per OB2 association. Therefore, our usual sanity check of 
confronting the color predictions obtained from the evolution models 
with those measured by standard photometry, does not work. Instead, 
we may get an estimate on the reddening and compute the true 
(dereddened) distance by assuming that the stellar parameters 
(most importantly \teff) are accurately estimated by our spectral 
and global analyses. The reddening is estimated with the help of 
two, partially overlapping methods. 

In the first method we compare the color index obtained from the 
evolution models with the one obtained from the 2MASS survey 
\citep{skrutskie:2006}. Here we use the ESO color system and 
transform the $J-K$ index\footnote{Unfortunately, direct 
estimation of $E(B-V)$ is not possible, due to the lack of accurate 
measurements in the $B$, $V$ filters for \hatcur{}.} from the 2MASS 
colors to the ESO colors as given by \cite{carpenter:2001}.   
This yields $(J-K)(\rm ESO)=$\hatcurCCesoJKmag\, for the observed 
color index. By using the isochrones value of 
$(M_{\rm J}-M_{\rm K})=$\hatcurISOJK\, and the relation of 
$E(B-V)=E(J-K_{\rm s})/0.505$ from \cite{bilir:2008}, we get 
$E(B-V)=0.291\pm0.078$.  

A more direct estimate of the reddening can be derived without the 
need of a color transformation by using the current near infrared 
color--temperature transformation of \cite{gonzalez:2009} based on 
the infrared flux method and 2MASS colors. We derive the dereddened 
color index by requiring their formula to give the same temperature 
as the one obtained from our SME analysis. At the abundance of 
\hatcurSMEiizfeh\, we get $(J-K_{\rm s})_0=0.395\pm0.023$, which 
yields $E(B-V)=0.313\pm0.075$, in a nice agreement with the one 
derived from the ESO-transformed 2MASS colors. Taking the arithmetic 
mean of the two values, we assign an overall reddening value of 
$E(B-V)=$\hatcurReddening\, to the system. 
  
With the above reddening, the apparent magnitude of 
$K_{\rm s}=$\hatcurCCtwomassKmag\, and the absolute $K_{\rm s}$~mag 
of $M_{\rm K}=3.13\pm0.08$, derived from the global analysis, we 
can compute the true (dereddened) distance of the system. Since the 
interstellar absorption is $A(K_{\rm s})=0.382E(B-V)=0.115\pm0.029$, 
the true (dereddened) distance modulus is $6.396\pm0.087$, corresponding 
to \hatcurXdist\,pc (as usual, the uncertainty excludes possible, 
difficult to quantify, systematics which may be present in the model 
isochrones, the color/temperature transformations, or the assumed 
interstellar absorption law).  

%
Here we examine the relation of \hatcur{} to the Per OB2 association. 
The average distance, radial velocity and proper motion of the 
association are, respectively, $318\pm27$\,pc, $+24\pm4$\,\kms\, and 
$15\pm1$\,\masyr\, \citep{belikov:2002, steenbrugge:2003}. 
From our own radial velocity measurements (see \refsecl{recspec}) 
we get that the average radial velocity of \hatcur{} is 
\hatcurDSgamma\kms\,. For the proper motion\footnote{The association 
is oriented in such a way that the projection of the true motion 
results in a relatively small proper motion \citep{steenbrugge:2003}. 
Therefore, because of the large errors, we do not make a comparison 
between the respective components of the proper motions.} 
we have $\sim19\pm6$\,\masyr\, from the UCAC3 catalog 
\citep[see][]{zacharias:2010}. We see that the kinematical parameters 
may justify a membership, but our target is considerably closer than 
the mean distance of the association, although, it is possible that 
the cluster extends somewhat closer than $200$~pc 
\citep[see Fig.~6 of][]{belikov:2002}. Additional evidence that 
\hatcur{} might lie within this association is that our analysis above 
predicts a reddening of $E(B-V)\sim0.30$, which is close to the typical 
value given by \cite{belikov:2002} (see their Fig.~4). The distance 
implies that \hatcur{} is situated in the outer region of Per OB2, 
nearer to our direction. This conclusion is supported also by the 
rather high value of the dust absorption of $E(B-V)\sim0.93$ given by 
the map of \cite{schlegel:1998}, versus the relative low value derived 
above from the analysis of the HIRES template spectrum.\footnote{We note 
however that the reddening map of \cite{schlegel:1998} may yield somewhat 
overestimated values at large reddenings 
\citep[cf.][]{arce:1999, schuster:2004}.} Thus, we conclude that 
\hatcur{} may well lie within the outskirts of the Per OB2 association. 
Nevertheless we can exclude the possibility that it is a physical member 
of the association with a high probability, based on its large age of 
\hatcurISOage{}\,Gyr. 

%
%
\newcommand{\hatcurisoshort}{YY}
\newcommand{\hatcurlumind}{\arstar}

\ifthenelse{\boolean{emulateapj}}{
    \begin{deluxetable}{lcl}
	\tablewidth{0pc}
}{
    \begin{deluxetable}{lcl}
	\tabletypesize{\scriptsize}
}
\tablecaption{
	Stellar parameters for \hatcur{}
	\label{tab:stellar}
}
\tablehead{
	\colhead{Parameter}	&
	\colhead{Value} &
	\colhead{Source}
}
\startdata
\sidehead{Spectroscopic parameters}
$\teffstar$ (K)\dotfill         &  \hatcurSMEteff   & SME\tablenotemark{a}\\
$\feh$ (dex)\dotfill            &  \hatcurSMEzfeh   & SME                 \\
$\vsini$ (\kms)\dotfill         &  \hatcurSMEvsin   & SME                 \\
$\vmac$ (\kms)\dotfill          &  \hatcurSMEvmac   & SME                 \\
$\vmic$ (\kms)\dotfill          &  \hatcurSMEvmic   & SME                 \\
$\gamma_{\rm RV}$ (\kms)\dotfill         &  \hatcurDSgamma       & DS      \\
\sidehead{Limb darkening and photometric parameters}
$a_{z}$\dotfill			&  \hatcurLBiz		& SME+Claret\tablenotemark{b}\\
$b_{z}$\dotfill			&  \hatcurLBiiz		& SME+Claret          \\
$a_{i}$\dotfill			&  \hatcurLBii		& SME+Claret\tablenotemark{b}  \\
$b_{i}$\dotfill			&  \hatcurLBiii		& SME+Claret          \\
$V$ (mag)\dotfill		&  \hatcurCCtassmv   & TASS                                      \\
$K$ (mag,ESO)\dotfill           &  \hatcurCCesoKmag  & 2MASS+Carpenter\tablenotemark{d}          \\
\sidehead{Derived parameters}
$\mstar$ ($\msun$)\dotfill      &  \hatcurISOmlong   & \hatcurisoshort+\hatcurlumind+SME\tablenotemark{c}\\
$\rstar$ ($\rsun$)\dotfill      &  \hatcurISOrlong   & \hatcurisoshort+\hatcurlumind+SME         \\
$\loggstar$ (cgs)\dotfill       &  \hatcurISOlogg    & \hatcurisoshort+\hatcurlumind+SME         \\
$\lstar$ ($\lsun$)\dotfill      &  \hatcurISOlum     & \hatcurisoshort+\hatcurlumind+SME         \\
$M_{\rm V}$ (mag)\dotfill       &  \hatcurISOmv      & \hatcurisoshort+\hatcurlumind+SME         \\
$M_{\rm K}$ (mag,ESO)\dotfill   &  \hatcurISOMK      & \hatcurisoshort+\hatcurlumind+SME         \\
$M_{\rm J}-M_{\rm K}$ (mag,ESO)\dotfill       &  \hatcurISOJK      & \hatcurisoshort+\hatcurlumind+SME         \\ 
Age (Gyr)\dotfill               &  \hatcurISOage     & \hatcurisoshort+\hatcurlumind+SME         \\
$E(B-V)$ (mag)\dotfill          &  \hatcurReddening  & \hatcurisoshort+\hatcurlumind+SME         \\
Distance (pc)\dotfill           &  \hatcurXdist      & \hatcurisoshort+\hatcurlumind+SME\tablenotemark{e}\\ [-1.5ex]
\enddata
\tablenotetext{a}{
	SME = ``Spectroscopy Made Easy'' package for analysis of
	high-resolution spectra \cite{valenti:1996}. These parameters
	depend primarily on SME, with a small dependence on the iterative
	analysis incorporating the isochrone fit and global modeling of
	the data, as described in the text.
}
\tablenotetext{b}{
	SME+Claret = Based on the SME analysis and tables of quadratic
        limb-darkening coefficients from \citet{claret:2004}.
}
\tablenotetext{c}{
	\hatcurisoshort+\hatcurlumind+SME = YY isochrones
	\citep{yi:2001}, \arstar\, luminosity indicator, and SME results.
}
\tablenotetext{d}{The 2MASS--ESO transformation is based on 
                  \cite{carpenter:2001}}. 
\tablenotetext{e}{Corrected for reddening}
\ifthenelse{\boolean{emulateapj}}{
    \end{deluxetable}
}{
    \end{deluxetable}
}

%
%
\subsection{Global modeling of the data}
\label{sec:globmod}

Our global analysis (involving the \lcs\, and radial velocity data, 
together with some feedback between the derived planetary and 
stellar parameters fully follows the methodology described in 
\citet{bakos:2010}. We summarize here only the most important steps
and refer to the paper mentioned and to \citet{pal:2009}. 
We simultaneously fit the HATNet light curve, the $z$ and $i$ band 
follow-up light curves, and the Keck RV observations. Our model for 
the follow-up light curves was based on \cite{mandel:2002}, while 
we used the limb-darkening free ``P1P3'' analytic approximation to 
this model \citetext{see \citealp{bakos:2010}} to fit the HATNet 
light curve. The limb darkening coefficients were computed from the 
tables of \citet{claret:2004} by using the stellar parameters derived 
from the spectral analysis (\refsecl{stelparam}). The transit shape was 
characterized by the relative planetary radius $p\equiv \rpl/\rstar$, 
the square of the impact parameter $b^2$, and the reciprocal of the 
half duration of the transit $\zrstar$. We also included a blending 
parameter $B_{\rm inst}$\footnote{$B_{\rm inst}$ is a scaling factor 
for the HATNet transit depth: 
$\delta_{\rm HATNet} = B_{\rm inst}(\rpl/\rstar)^2$.} 
for the HATNet data to take into consideration the much lower 
resolution of the HATNet images than those taken with KeplerCam.

As far as the RV data are concerned, we adjusted four parameters, 
corresponding to an unperturbed Keplerian orbit: the overall zero 
point shift $\gamma_{\rm rel}$ of the velocity values, the semiamplitude 
$K$ and Lagrangian orbital elements $(k,h)=e\times(\cos\omega,\sin\omega)$. 
The moment of the center of the transit $T_c$ is fixed by the photometric 
data and determined by the epoch numbers and by two individual events 
(of course, assuming strict periodicity)\footnote{We prefer 
using two epochs rather than an epoch and the period, because we would 
like to derive quantities that are as weakly correlated as possible -- 
see \citet{bakos:2007} and \citet{pal:2008}.}. We assigned the transit 
number $N_{tr} =0$ to the first high quality follow-up \lc\ gathered on 
2007 December 18. The adjusted moments of transit centers in the fit were 
the first transit center observed with HATNet at $T_{c,-74}$, and the 
last high quality transit center observed with the \flwof\ telescope,
$T_{c,+28}$. Other events were then defined by using these two epochs 
and the corresponding $N_{tr}$ transit numbers. 

Systematics in the HATNet data were treated by EPD/TFA-filtering prior 
to the global analysis. For the FLWO data we employed a joint iterative 
filtering of the above type together with the search for the best fitting 
physical model. The five EPD parameters were the hour angle, the square 
of the hour angle, and the stellar profile parameters (equivalent to FWHM, 
elongation and position angle). The EPD procedure was performed in ``local'' 
mode with separate EPD coefficients defined for each night. The five 
EPD parameters were augmented by the out-of-transit magnitude 
of the individual events.  The TFA filtering was performed in ``global'' 
mode using the same set of stars and TFA coefficients for all nights. 
We chose 20 template stars that had good quality measurements for all 
nights and on all frames, implying an additional 20 parameters in the fit. 
The number of fitted parameters were the following: 
$4\times6=24$ EPD, $20$ TFA, $3$, related to instrumental configuration 
(the blend factor $B_{\rm inst}$ and the out-of-transit magnitude, 
$M_{\rm 0,HATNet}$ of HATNet and the relative RV zero-point $\gamma_{\rm rel}$) 
and the $8$ physical model parameters (transit times $T_{c,-74}$, $T_{c,+28}$ 
-- for transit epoch and period, $\rpl/\rstar$, $b^2$, $\zrstar$, $K$, 
$k=e\cos\omega$, $h=e\sin\omega$). The total number of fitted parameter 
is $55$, which is well below the number of available data points ($1786$ 
photometric followup and $23$ radial velocity data). For additional 
technical details see \citet{bakos:2010} and \citet{pal:2009}. We note 
only that the errors were obtained through Monte-Carlo simulations as 
described, e.g., by \cite{ford:2006}. 

The final solution for the relevant planetary parameters are summarized in 
\reftabl{planetparam}.\footnote{Some auxiliary parameters (not listed 
in \reftabl{planetparam}) are:
$T_{\mathrm{c},-74}=\hatcurLCTA$~(BJD),
$T_{\mathrm{c},+28}=\hatcurLCTB$~(BJD),
$\gamma_{\rm rel}=\hatcurRVgamma$\,\ms\,. This average velocity for the 
Keck RVs is merely a zero point value and does \emph{not} correspond to the 
true center of mass RV of the system (for this see the corresponding DS value 
in \reftabl{stellar}), $B_{instr}=\hatcurLCiblend$.}

Also listed is the ``RV jitter'', which is a component of assumed 
astrophysical noise intrinsic to the star that we add in quadrature to 
the RV measurement uncertainties in order to have $\chi^{2}/{\rm dof} = 1$ 
from the RV data for the global fit (see also \refsecl{hispec} for further 
discussion of the jitter). For possible future observation of occultation 
events we computed the eclipse parameters. Since the transit happens near 
apastron, the occultation occurs with certainty.

%
%
\ifthenelse{\boolean{emulateapj}}{
    \begin{deluxetable}{lc}
	\tablewidth{0pc}
}{
    \begin{deluxetable}{lc}
	\tabletypesize{\scriptsize}
}
\tablecaption{Orbital and planetary parameters of \hatcur\,{\rm b} 
\label{tab:planetparam}}
\tablehead{
	\colhead{~~~~~~~~~~~~~~~Parameter~~~~~~~~~~~~~~~} &
	\colhead{Value}
}
\startdata
\sidehead{\Lc{} parameters}
~~~$P$ (days)             \dotfill    & $\hatcurLCP$              \\
~~~$T_c$ (${\rm BJD}$)    
      \tablenotemark{a}   \dotfill    & $\hatcurLCT$              \\

~~~$T_{14}$ (days)
      \tablenotemark{b}   \dotfill    & $\hatcurLCdur$            \\
~~~$T_{12} = T_{34}$ (days)
    \tablenotemark{b}     \dotfill    & $\hatcurLCingdur$         \\
~~~$\arstar$              \dotfill    & $\hatcurPPar$             \\
~~~$\zrstar$              \dotfill    & $\hatcurLCzeta$           \\
~~~$\rpl/\rstar$          \dotfill    & $\hatcurLCrprstar$        \\
~~~$b^2$                  \dotfill    & $\hatcurLCbsq$            \\
~~~$b \equiv a \cos i/\rstar$
                          \dotfill    & $\hatcurLCimp$            \\
~~~$i$ (deg)              \dotfill    & $\hatcurPPi$ \phn         \\

\sidehead{RV parameters}
~~~$K$ (\ms)              \dotfill    & $\hatcurRVK$              \\
%
~~~$k_{\rm RV}$\tablenotemark{c} 
                          \dotfill    & $\hatcurRVk$              \\
~~~$h_{\rm RV}$\tablenotemark{c}
                          \dotfill    & $\hatcurRVh$              \\
~~~$e$                    \dotfill    & $\hatcurRVeccen$          \\
%
~~~$\omega$                  \dotfill    & $\hatcurRVomega^\circ$ \\
~~~$\sigma_{\rm RV}$(fit) (\ms) \dotfill    & \hatcurRVrms           \\
~~~RV jitter (\ms)           \dotfill    & \hatcurRVjitter        \\

\sidehead{Secondary eclipse parameters}
~~~$T_s$ (BJD)            \dotfill    & $\hatcurXsecondary$       \\
~~~$T_{s,14}$ (days)      \dotfill    & $\hatcurXsecdur$          \\
~~~$T_{s,12}$ (days)      \dotfill    & $\hatcurXsecingdur$       \\

\sidehead{Planetary parameters}
~~~$\mpl$ ($\mjup$)       \dotfill    & $\hatcurPPmlong$          \\
~~~$\rpl$ ($\rjup$)       \dotfill    & $\hatcurPPrlong$          \\
~~~$C(\mpl,\rpl)$
    \tablenotemark{d}     \dotfill    & $\hatcurPPmrcorr$         \\
~~~$\rhopl$ (\gcmc)       \dotfill    & $\hatcurPPrho$            \\
~~~$a$ (AU)               \dotfill    & $\hatcurPParel$           \\
~~~$\log g_p$ (cgs)       \dotfill    & $\hatcurPPlogg$           \\
~~~$T_{\rm eq}$ (K)       \dotfill    & $\hatcurPPteff$           \\
~~~$\Theta$\tablenotemark{e}               \dotfill    & $\hatcurPPtheta$          \\
%
~~~$F_{per}$ ($10^{\hatcurPPfluxperidim}$\ergscmsq) \tablenotemark{f}
                          \dotfill    & $\hatcurPPfluxperi$      \\
~~~$F_{ap}$  ($10^{\hatcurPPfluxapdim}$\ergscmsq) \tablenotemark{f} 
                          \dotfill    & $\hatcurPPfluxap$        \\
~~~$\langle F \rangle$ ($10^{\hatcurPPfluxavgdim}$\ergscmsq) \tablenotemark{f}
                          \dotfill    & $\hatcurPPfluxavg$        \\ [-1.5ex]
\enddata
\tablenotetext{a}{
	All Julian dates in this paper are based on the UTC scale. For
	applications requiring precisions of the order of a second, such as the
	study of transit timing variations, the use of Terrestrial Time is
	recommended (see, e.g.,\cite{torres:2010}).
}
\tablenotetext{b}{
	\ensuremath{T_{14}}: total transit duration, time
	between first to last contacts;
	\ensuremath{T_{12}=T_{34}}: ingress/egress duration,
	time between first and second, or third and fourth contacts.
}
\tablenotetext{c}{
	Lagrangian orbital parameters derived from the global modeling, and
	primarily determined by the RV data.
}
\tablenotetext{d}{
	Correlation coefficient between the planetary mass \mpl\ and radius
	\rpl.
}
\tablenotetext{e}{
	The Safronov number is given by $\Theta = \frac{1}{2}(V_{\rm
	esc}/V_{\rm orb})^2 = (a/\rpl)(\mpl / \mstar )$
	\citep[see][]{hansen:2007}.
}
\tablenotetext{f}{
	Incoming flux per unit surface area.
}
\ifthenelse{\boolean{emulateapj}}{
    \end{deluxetable}
}{
    \end{deluxetable}
}


%
%
\section{Discussion}
\label{sec:discussion}

The transiting extrasolar planet (TEP) \hatcurb{} reported in this paper 
is among the few with orbital periods greater than 
$8$~days.\footnote{Currently, there are no TEPs between $5.6$ and $8$~days, 
and, together with \hatcurb{}, there are only four TEPs with periods 
greater than $10$~days. The periods are more densely distributed below 
$5$~days.} We summarize the properties of the currently known long-period 
TEPs in \reftabl{ltep}. 

%
%
\ifthenelse{\boolean{emulateapj}}{
    \begin{deluxetable*}{lccrrcccrcccccr}
}{
    \begin{deluxetable}{lccrrcccrcccccr}
	\rotate
}
\tabletypesize{\scriptsize}

\tablewidth{0pc}
\tablecaption{Long-period TEPs\tablenotemark{a}
\label{tab:ltep}}
\tablehead{
	\colhead{Name} & 
	\colhead{$\mpl$} & 
	\colhead{$\rpl$} & 
	\colhead{$M_{\rm core}$\tablenotemark{b}} & 
	\colhead{$T_{\rm eq}$\tablenotemark{c}} & 
	\colhead{Period} &
	\colhead{a} & 
	\colhead{e} & 
	\colhead{$T_{14}$} &  
	\colhead{Star(\teff)} & 	
	\colhead{[Fe/H]} & 	
	\colhead{Age\tablenotemark{e}} & 	
	\colhead{$\mstar$} & 	
	\colhead{$\rstar$} &
	\colhead{V} 	\\
	\colhead{} & 
	\colhead{[$\mjup$]} & 
	\colhead{[$\rjup$]} & 
	\colhead{[\mearth]} & 
	\colhead{[K]} & 
	\colhead{[d]} &
	\colhead{[AU]} & 
	\colhead{} & 
	\colhead{[h]} &  
	\colhead{[K]} & 	
	\colhead{[dex]} & 	
	\colhead{[Gyr]} & 	
	\colhead{[$\msun$]} & 	
	\colhead{[$\rsun$]} &
	\colhead{[mag]}
}
\startdata
WASP-8b      & 2.25    & 1.05  & 52  &  940 & 8.15872   & 0.080 & 0.31  & 3.5 & G6 
             (5600)  & $+0.17$ & 4.0 & 1.03 & 0.95      &  9.9 \\   
CoRoT-6b     & 2.96    & 1.17  & 0 & 1020 & 8.88659   & 0.086 & 0.10\tablenotemark{d}  & 3.8 & F5 
             (6090)  & $-0.20$ & 3.0 & 1.06 & 1.03      & 13.9 \\
CoRoT-4b     & 0.72    & 1.19  & 0 & 1070 & 9.20205   & 0.090 & 0.00  & 4.4 & F0 
             (6190)  & $+0.00$ & 1.0 & 1.10 & 1.15      & 13.7 \\
HAT-P-15b    & \hatcurPPmshort 
             & \hatcurPPrshort
	     & $10$ 
	     & 910 
	     & \hatcurLCPshort 
	     & $0.096$ 
	     & $0.19$ 
	     & $5.5$ 
	     & G5 (5568) 
	     & \hatcurSMEiizfehshort
	     & \hatcurISOageshort 
	     & \hatcurISOmshort
	     & \hatcurISOrshort  
	     & 12.2 \\
HD~17156b    &	 3.21  & 1.02  & 136  & 880  &	21.2169   & 0.162 & 0.68  &  3.2 & G0 
             (6079)  & $+0.24$ & 3.1  & 1.24 & 1.45      &  8.2 \\
CoRoT-9b     &	 0.84  & 1.05  & 18   & 410  &	95.2738   & 0.407 & 0.11  &  8.1 & G3 
             (5625)  & $-0.01$ & 3.0  & 0.99 & 0.94      & 13.7 \\  	
HD~80606b    &	 4.08  & 0.98  & 195  & 400  &	111.436   & 0.455 & 0.93  & 11.9 & G5 
             (5574)  & $+0.43$ & 7.6  & 1.01 & 1.01      & 8.9 \\ [-1.5ex]
\enddata
\tablenotetext{a}{All parameters are from {\em http://exoplanet.eu/}, 
except for those of WASP-8b \cite[see][]{queloz:2010} and HD~80606b 
\citetext{see \citealp{moutou:2009} (\teff) and \citealp{hebrard:2010}}.}
\tablenotetext{b}{Based on the models of \cite{fortney:2007}; zero core 
mass indicates that the planet has larger radius than predicted by the 
models at zero core mass.}
\tablenotetext{c}{Computed at a distance given by the semi-major axis.}
\tablenotetext{d}{Upper limit.}
\tablenotetext{e}{Ages have been taken either from `exoplanet.eu' or 
from currently published papers. For CoRoT-6b we adopted an intermediate 
value from the range of $2.5$--$4.0$~Gyr listed by \cite{fridlund:2010}. 
For CoRoT-9b \cite{deeg:2010} give a wide range of $0.2$--$8$~Gyr, and 
we adopted a plausible intermediate-to-low age.}
\ifthenelse{\boolean{emulateapj}}{
    \end{deluxetable*}
}{
    \end{deluxetable}
}

\hatcurb{} is the longest period TEP discovered by ground-based 
photometric surveys. With a duration of $5.5$~hours, the transit of 
\hatcurb{} is at the limit in length that can be pursued from a single 
site ground-based observation (the only planet with a longer period 
than \hatcurb{} that can still be followed up from a single site is 
HD~17156b). The possibility of performing single-site full followup 
observations is a valuable property, since systematics are more difficult 
to handle when low signal-to-noise ratio data from different sites 
are combined. The orbit is eccentric with a high significance. Except 
for CoRoT-4b (and perhaps for CoRoT-6b), all of the 7 long-period 
planets have this property. Most of the planet masses are greater 
than $2\mjup$. The host star masses are lower than $\sim1.2$\msun\,, 
in agreement with the upper mass limit of $\sim1.4$\msun\, of the 
extrasolar planets situated in the `period valley' between 
$\sim0.1$~AU and $\sim1$~AU \citep{bowler:2010}.  

The relative depletion of planets in the $0.1$--$1$~AU region was 
recognized a few years ago when the number of extrasolar planets 
grew to the level that made more reliable statistical investigations 
possible \citep{johnson:2007}. All host stars with semi-major axes 
smaller than $\sim1$~AU have masses lower than $\sim1.4$\msun\,. 
Above $\sim1$~AU there are several stars with $M\sim0.5$--$2.5$\msun\, 
\citep{bowler:2010}. The distribution of the periods (or semi-major 
axes) of all known extrasolar planets is bimodal with a high significance 
(see upper panel of \reffigl{pdistr}). Although the sample of multiple 
systems is not very extensive, it is interesting to note that the 
distribution becomes more uniform when considering only these systems 
\cite{wright:2009}. Although some aspects of these properties can 
be explained by engulfment of planets in evolved giants 
\citep{villaver:2009}, early stage of planet formation should 
play more important role. According to the models considering 
the formation phase 
\citetext{\citealp{burkert:2007} and \citealp{currie:2009}, see 
also \citealp{kretke:2009} for the role of magnetorotational 
instability in the protoplanetary disk} the bimodal distribution  
is attributed to the stellar-mass-dependent lifetime of the 
protoplanetary gas disk. For higher mass stars, the gas-depletion 
time scale is shorter due to accretion and increased EUV radiation, 
and therefore planets may stop inward migration too early and get 
stranded on orbits at high semi-major axes. For stellar masses 
$\la1.2$\msun\, planets do migrate. The rate of migration is 
determined by the ratio of the viscous and depletion time 
scales.\footnote{With a careful filtering of the currently available 
data we found that the valley exist at all host star masses, albeit 
for stars with $\ga1.2$\msun\, the effect is indeed much stronger.}   

\begin{figure}[!ht]
\plotone{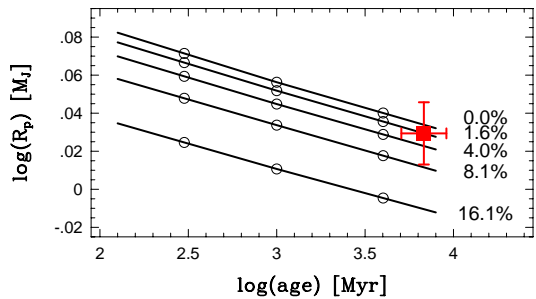}
\caption{
         Position of \hatcurb{} on the age --- planet radius diagram. 
	 The models of \cite{fortney:2007} have been interpolated to the 
	 derived mass and solar equivalent semi-major axis of \hatcur{}. 
	 Open circles are the tabulated/interpolated values from 
	 \cite{fortney:2007}, continuous lines have been obtained by 
	 linear interpolation. Lines are labeled by the core mass 
	 (relative to the total planet mass).  
}
\label{fig:fortney-model}
\end{figure}

We investigated the agreement between the derived planetary mass and 
radius and predictions from current theoretical models. We interpolated 
the models of \cite{fortney:2007} to the observed planet mass of 
$\hatcurPPmlong$~$\mjup$ and semi-major axis of equivalent solar 
irradiance (which is the same as the observed semi-major axis in this 
case). The iso-core lines in the $\log(age)$--$\log(R_{\rm p})$ plane 
are shown in \reffigl{fortney-model}. Although the errors are large, 
and the derived core mass (especially at the low end) depends very 
sensitively on the stellar/planet parameters, it seems likely that 
\hatcurb{} is a H/He-dominated gas giant planet with a core mass most 
likely about $10$\mearth\,, but which could range from zero to perhaps 
$50$\mearth\,.

\hatcurb{} does not exhibit the ``radius anomaly'', where -- as is 
frequently the case for hot Jupiters -- the planetary radius is 
significantly larger than predicted by models such as that of 
\cite{fortney:2007}. Furthermore, we find that for a hypothetical 
planet with the same mass and radius as those of \hatcurb{}, the  
derived core mass is only a weak function of the period (semi-major axis) 
between $\sim10$ and $1000$~days (it varies between $0$ and $10$\mearth\,). 
However, below $10$~days the core mass increases up to $60$\mearth\, 
(at $1$~day), and starts especially being sensitive below $4$~days. 
These all show that \hatcurb{} resides in the period regime where 
standard planet structure models (i.e., the ones already tested on 
Solar System planets) are satisfactory, without resorting to additional 
effects (e.g., extra heat source, tidal heating, enhanced atmospheric 
opacities, etc., see \citealp{miller:2009}) required by many of the 
hot Jupiters. 

In \reftabl{ltep} we also added the derived core masses for the other 
long-period TEPs. We see that two of the shorter period members still 
exhibit some radius anomalies. For CoRoT-4b and CoRoT-6b, tidal 
heating \citep{miller:2009} is unlikely, since they both have close 
to circular orbits. On the other hand, considering the possible 
ambiguities in the observed absolute planetary radii, it is possible 
that future followup observation will result in downward radius 
corrections of $3$--$5$\% necessary to solve the apparent anomaly 
for these planets. This level of radius correction is not uncommon 
in the followup works of increasing accuracy 
\citetext{e.g., the case of HAT-P-1b -- \citealp{winn:2007} or that 
of HD~80606b -- \citealp{fossey:2009} versus \citealp{hebrard:2010}, 
and especially WASP-8b, with $10$\% radius decrease in the very recent 
analysis of \citealp{queloz:2010}}. 

\begin{figure}[!ht]
\plotone{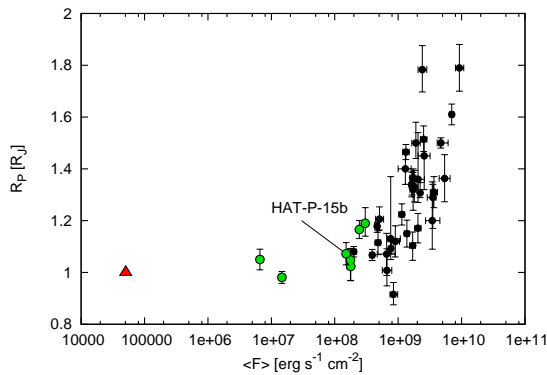}
\caption{
     Incident stellar flux versus planet radius for TEPs of 
	 $0.7\la M_{\rm p}/M_{\rm J}\la 5.0$. The long-period 
	 planets of \reftabl{ltep} are shown by whiter shade (green), 
	 the triangle at the left corner shows the position of Jupiter.
}
\label{fig:flux-radius}
\end{figure}

It is also useful to compare the overall dependence of the planetary 
radius on the incident stellar flux. \reffigl{flux-radius} shows this 
dependence for TEPs in the mass range of $(0.7$--$5.0)$\mjup. Although 
the stellar fluxes for the long-period TEPs are still $2$--$3$ orders 
of magnitude larger than that of Jupiter, it seems that this is already 
enough to cause a substantial decrease in the planetary radius, and 
bring down the otherwise inflated values to `more normal' ones, 
corresponding to that of the Jupiter. We note that this happens in a 
quite wide mass range.  

%
%

%
%

With V=$12.2$~mag, \hatcur{} is still bright enough to be the subject of 
ground- and space-based observations. At a distance of $a=\hatcurPParel$~AU 
from its host star, \hatcurb{} allows us to investigate the properties of a 
colder planet, not influenced by the complicated hydrodynamical and radiative 
effects which may apply to canonical hot Jupiters (see, e.g., the discussion 
of the core mass insensitivity above). At the same time, the period of 
$\hatcurLCPvshort$~days is still on a relatively short time scale, 
allowing frequent followup observations. The feasibility of possible 
follow-up observations is briefly summarized below. 
\begin{itemize}
\item
{\em Search for other planets.}
Long-term radial velocity (RV) and transit timing monitoring is 
justified due to the signature of excess RV scatter that is unlikely 
to be due to stellar activity (see \refsecl{hispec}). Although we have 
made a very thorough search both in the RV and in the photometric data 
for a trace of a second planet, we have not found anything convincing. 
Due to the relatively small value of \hatcurRVjitter\ms\, of the ``jitter'', 
the mass of the possible hidden companion is estimated in the sub-Neptune 
regime (except if there is a planet in, e.g., a 2:1 resonance, hiding 
in the eccentric solution -- see \citealp{anglada:2010}). We note that 
here, the eccentricity of the orbit of \hatcurb{} is a less forcing 
argument for searching for a putative second planet, because at the 
semi-major axis of \hatcurb{} the circularization time scale due to 
tidal dissipation may be on the order of several Gyrs \citep{matsumura:2008}.     
\item
{\em Inclination of the orbital and stellar spin axes.} 
The measurement of the Rossiter-McLaughlin (R-M) effect for \hatcur{} is 
somewhat challenging, due to its relatively low projected rotation rate of 
\vsini$=$\hatcurSMEiivsin\,. The expected size of the effect is $\sim20$\ms\,, 
similar to the one currently measured for WASP-8b with the same \vsini\, 
\citep{queloz:2010}. Three of the seven long-period planets listed in 
\reftabl{ltep} have yielded R-M data sufficient to compute the projected 
inclination of the stellar rotational and orbital axes. HD~17156 is aligned 
\citep{narita:2009}, while HD~80606b is tilted \citep{hebrard:2010}, and 
WASP-8b \citep{queloz:2010} is apparently strongly tilted, suggesting a 
retrograde orbit. It is clear that measuring the R-M effect of long-period  
planets in more evolved systems such as \hatcur{} is very important, due 
to the expected relaxation of the dynamics to a final equilibrium state. 

\item
{\em Thermal emission.} 
With a proper choice of the infrared waveband and repeated observations, 
the detection of the occultation of \hatcurb{} by ground-based instruments 
may be feasible. Assuming blackbody radiation, at a fixed wavelength the 
occultation depth is equal to 
$\delta_{\rm occ}=(R_{\rm p}/R_{\rm star})^2(T_{\rm pl}/T_{\rm star})$ 
\citep{winn:2010}. Assuming $T_{\rm eq}=T_{\rm pl}$, at the moment of 
occultation (which is the periastron in the case of \hatcurb{}) we have 
$T_{\rm pl}\sim1000$~K. It follows then that $\delta_{\rm occ}\sim0.002$, 
which is measurable, especially if we consider that the occultation lasts 
for $\sim4$~hours; short-enough to cover the event in one night and 
long-enough to gather sufficient number of data points (especially in 
the case of repeated observations) to reach the level of detection of 
several $\sigma$.  
\item
{\em Atmospheric absorption}. 
Measuring the small change in the transit depth due to the varying 
absorption level in different wavelengths (i.e., performing transmission 
spectroscopy) is even more challenging than detecting the occultation 
event. This is because the change in the transit depth is proportional 
to the scale-height $H$, which is proportional to the ratio of the local 
temperature to the gravity $T/g$ \citep{winn:2010}. Since \hatcurb{} 
is relatively cold and has a higher gravity than `standard' hot Jupiters, 
it is expected that the signal will be about $25$\% of what is usually 
measured (assuming `standard' hot Jupiter $T$ and $g$ of $2000$~K and 
$20$\mss\,). Considering that the relative radius variation even for 
short-period targets like HD~189733 is of the order of $0.1$\% 
\citep{sing:2009}, measuring absorption features in colder planets such 
as \hatcurb{} would indeed be very difficult. 
\end{itemize}

With the continuing work of ground-based wide-field photometric surveys, 
we expect to discover further TEPs in the $10$--$30$~day period regime. 
Since these will be mostly bright ($V\la13$~mag) targets with orbital 
periods still considered to be relatively short, a wide range of followup 
works will be feasible. Therefore, these planets will fill the gap between 
the classical hot Jupiters and the long-period, but likely much fainter 
planetary systems (more akin to the Solar System) to be discovered by 
ongoing and future space projects.


\acknowledgements 

HATNet operations have been funded by NASA grants NNG04GN74G,
NNX08AF23G and SAO IR\&D grants. Work of G.\'A.B.~and J.~Johnson were
supported by the Postdoctoral Fellowship of the NSF Astronomy and
Astrophysics Program (AST-0702843 and AST-0702821, respectively). GT
acknowledges partial support from NASA grant NNX09AF59G. We acknowledge
partial support also from the Kepler Mission under NASA Cooperative
Agreement NCC2-1390 (D.W.L., PI). G.K.~thanks the Hungarian Scientific
Research Foundation (OTKA) for support through grant K-81373. This
research has made use of Keck telescope time granted through NOAO
(programs A285Hr and A146Hr) and NASA (programs N128Hr, N145Hr and 
N018Hr).



\end{document}